%% file: conference_101719.tex
\documentclass[conference]{IEEEtran}
\IEEEoverridecommandlockouts
\usepackage{cite}
\usepackage{amsmath,amssymb,amsfonts}
\usepackage{algorithmic}
\usepackage{graphicx}
\usepackage{multirow}
\usepackage{textcomp}
\usepackage{xcolor}
\usepackage{array}
\usepackage{tabularx}
\newcolumntype{L}[1]{>{\raggedright\arraybackslash}m{#1}}
\newcolumntype{C}[1]{>{\centering\arraybackslash}m{#1}}
\newcolumntype{R}[1]{>{\raggedleft\arraybackslash}m{#1}}
\def\BibTeX{{\rm B\kern-.05em{\sc i\kern-.025em b}\kern-.08em
    T\kern-.1667em\lower.7ex\hbox{E}\kern-.125emX}}
\begin{document}
\newcommand{\hide}[1]{}
\newcommand{\datakop}{LTW1\,}
\newcommand{\dataeighteighty}{STW1\,}
\newcommand{\datamltest}{LTW2\,}
\newcommand{\ourmethod}{CAAD\,}
\newcommand{\ourmethodUQ}{CAAD-UQ\,}
\newcommand{\ourframework}{CAAD-EF\,}
\newcommand{\nikhilc}[1]{\textcolor{red}{#1}}
\newcommand{\fortim}[1]{\textcolor{blue}{Tim: #1}}
\newcommand{\forgopi}[1]{\textcolor{orange}{Gopi: #1}}
\title{Detecting Irregular Network Activity with\\ Adversarial Learning and Expert Feedback}

\author{\IEEEauthorblockN{Gopikrishna Rathinavel}
\IEEEauthorblockA{\textit{Virginia Tech} \\
Blacksburg, VA\\
rgopikrishna@vt.edu}
\and
\IEEEauthorblockN{Nikhil Muralidhar}
\IEEEauthorblockA{\textit{Stevens Institute of Technology} \\
Hoboken, NJ\\
nmurali1@stevens.edu}
\and
\IEEEauthorblockN{Timothy O'Shea}
\IEEEauthorblockA{\textit{DeepSig Inc \& Virginia Tech} \\
Arlington, VA\\
tim@deepsig.io}
\and
\IEEEauthorblockN{Naren Ramakrishnan}
\IEEEauthorblockA{\textit{Virginia Tech} \\
Arlington, VA\\
naren@cs.vt.edu}

}

\maketitle

\begin{abstract}
Anomaly detection is a ubiquitous and challenging task, relevant across many disciplines. With the vital role communication networks play in our daily lives, the security of these networks is imperative for the smooth functioning of society. To this end, we propose a novel self-supervised deep learning framework \ourmethod for anomaly detection in wireless communication systems. Specifically, \ourmethod employs contrastive learning in an adversarial setup to learn effective representations of normal and anomalous behavior in wireless networks. We conduct rigorous performance comparisons of \ourmethod with several state-of-the-art anomaly detection techniques and verify that \ourmethod yields a mean performance improvement of \textbf{92.84\%}. Additionally, we also augment \ourmethod enabling it to systematically incorporate expert feedback through a novel contrastive learning feedback loop to improve the learned representations and thereby reduce prediction uncertainty (\ourframework). We view \ourframework as a novel, holistic, and widely applicable solution to anomaly detection. Our source code and data are available online\footnote{https://github.com/rgopikrishna-vt/CAAD}
\end{abstract}


\section{Introduction}\label{sec:introduction}
\input{sections/introduction}

\section{Related Work}\label{sec:related_work}
\input{sections/related_work}

\section{Background}\label{sec:background}
\input{sections/background}
\section{Problem Formulation}\label{sec:problem_formulation}
\input{sections/problem_formulation}
\section{Experimental Setup}\label{sec:experimental_setup}
\input{sections/experimental_setup}
\section{Results and discussion}\label{sec:results}

\input{sections/results_and_discussion}
\section{Conclusion}
\input{sections/conclusion}
\bibliographystyle{IEEEtran}
\bibliography{bibfile}
\appendix
\section{Appendix}\label{sec:appendix}
\input{sections/appendix}

\end{document}

%% file: sections/introduction.tex
Wireless communications systems form an essential component of cyber-physical systems in urban environments along with the electric grid and the transportation network. These wireless communication systems and networks enable us to access the internet, and connect with others remotely, thereby serving as a vital means for human interaction.  Further, they connect hundreds or thousands of sensors, applications, industrial networks, critical communications systems, and other infrastructure.  Hence, state monitoring and detection of irregular activity in wireless networks are essential to ensuring robust and resilient system operational capabilities.

The \emph{electromagnetic spectrum} (simply referred to as `the spectrum') is the information highway through which most modern forms of electronic communication occur. Parts of the spectrum are grouped into `bands' (based on the wavelength) which can be thought of as analogous to lanes on the highway. Specific regions (i.e., lanes) of the spectrum are reserved for specific types of communication (e.g., radio communication, broadcast television) based on frequency. The entire spectrum ranges from 3Hz-300EHz and the typical range used for \emph{wireless communication} today is 30Khz-28GHz. 

\begin{figure}[!t]
    \vspace{1ex}
    \centering
    \includegraphics[width=0.48\textwidth]{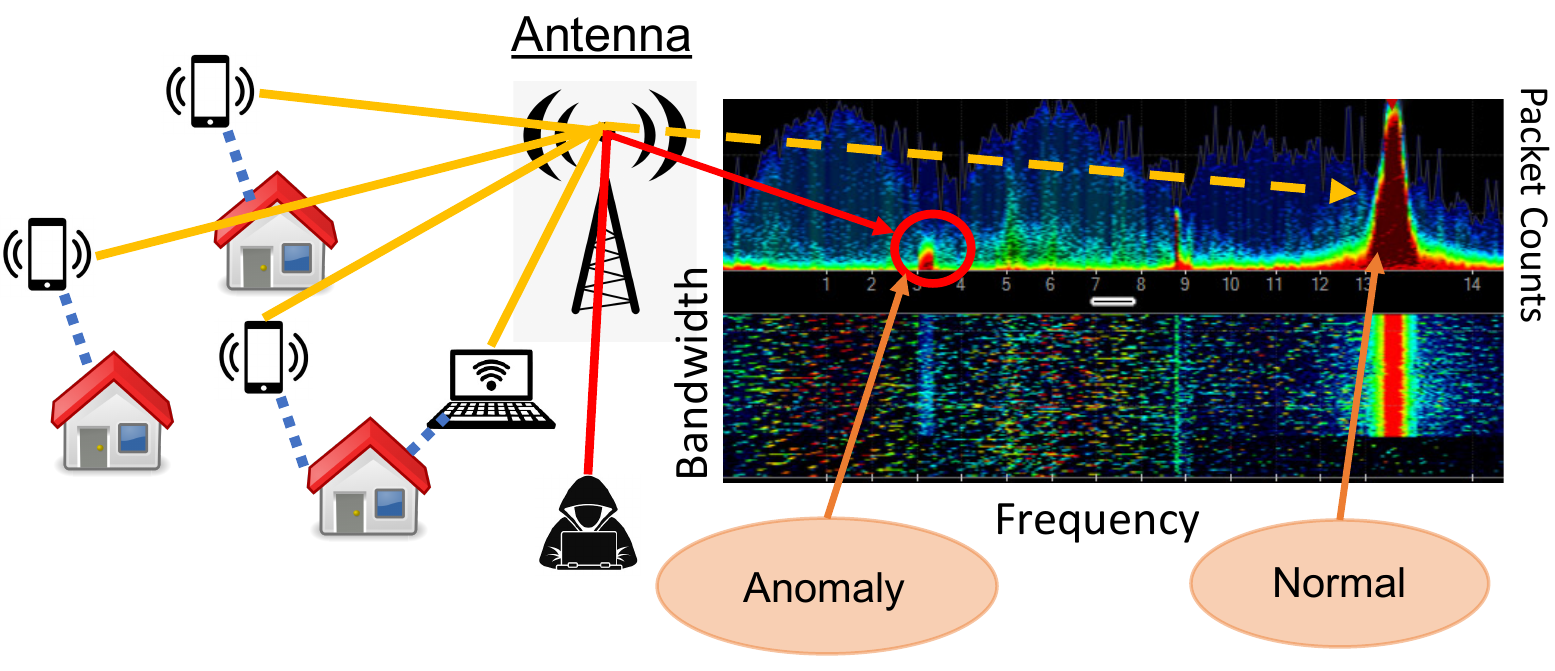}
    \caption{Irregular Activity in Wireless Communication Systems. \cite{metageek}}
    \label{fig:introfig}
    \vspace{-3ex}
\end{figure}

\emph{Spectrum access activity} in wireless systems carries rich information which can indicate underlying activity of physical device presence, activity and behaviors corresponding to security threats and intrusions, jamming attempts, device malfunctioning, interference, illicit transmissions, and a host of other activities (see Fig.~\ref{fig:introfig}). Data corresponding to spectrum access activity information has been explored in wireless intrusion detection systems (WIDS) in a very limited context and most of the systems in use today for detecting anomalous network activity, are highly application specific and focus on specialized feature engineering, detector engineering, and signal-specific digital signal processing (DSP) engineering.\hide{and has been used as well in cellular-specific interference and coverage mapping tools, as well as in numerous other wireless monitoring systems which are often designed to monitor and analyze traffic and phenomenon for a specific type of radio system. Examples of these include binary access prediction within the context of simple dynamic spectrum access (DSA) systems~\cite{eltom2018statistical} and protocol-dependent feature-based prediction within the context of one technology such as LTE cellular access~\cite{7811244}.} Such systems are not generalizable, are highly sensitive to minor variations in system characteristics and are costly to maintain due to the requirements of rich feature engineering.

Hence in this work, we have developed a generic and powerful unsupervised anomaly detection framework and demonstrated its prowess in the context of wireless network anomalies. Specifically, we propose a novel solution to anomaly detection (AD), \emph{Contrastive Adversarial Anomaly Detection} (\ourmethod) which applies contrastive learning (CL) in an adversarial setup. We also augment CAAD with the ability to incorporate expert feedback (EF) to improve the quality of its learned representations for AD. We call this model \ourframework (\ourmethod with expert feedback). To the best of our knowledge, we are the first to propose such a powerful yet flexible AD framework that applies contrastive learning paradigms in an adversarial setup with the ability to incorporate expert feedback via contrastive learning to improve its learned representations and reduce prediction uncertainty.

\hide{with the ability to improve representation learning capability by fine-tuning representations based on expert feedback employing novel contrastive learning techniques.

Opening up all RF frequency bands to such access modeling, AD, and change detection is an unprecedented capability that offers to help optimize and secure future wireless IoT systems, cellular systems, WLAN systems, and a wide range of emergency communications, telemetry, broadcast, and other systems which we rely on communicating through and sharing the spectrum every day.} \hide{By reducing the data-flood of high rate raw sensor data to a smaller meta-data stream, and by applying novel time series modeling approaches to predicting access within this feature space, we demonstrate that the same models can be effectively applied to several different bands (namely ISM and GMR bands in this work) which employ completely different technology types, access strategies, and time-scales, and both converge to effective band-access prediction models using the same class of data-driven feature extraction and sequence modeling models.}
\par\noindent
Our contributions are as follows:
\par\noindent
$\bullet$ We propose \ourmethod, a novel method for AD which utilizes contrastive learning and generative adversarial networks (GAN). We demonstrate that our proposed model is able to significantly outperform state-of-the-art (SOTA) models on AD in wireless networks and standard datasets. To the best of our knowledge, \ourmethod is the first model to use a combination of CL and adversarial learning for AD. 
\par\noindent
$\bullet$ We propose \ourframework, which is another novel model supplemental to \ourmethod,  which further enables us to incorporate expert feedback via contrastive learning and uncertainty quantification using Monte Carlo dropouts. To the best of our knowledge, our framework is the first successful undertaking to utilize contrastive learning to incorporate expert feedback.
\par\noindent
$\bullet$ Finally, we highlight the importance of various facets of~\ourframework through rigorous qualitative, quantitative, and ablation analyses.

\hide{We believe this is an important step towards widely applicable RF monitoring and change detection systems, which can provide a broad scope of analytics and high-level event detection to provide insight into activity occurring on numerous RF bands across numerous radio technology devices.  As our global spectrum policy turns increasingly towards dynamic and adaptive spectrum access policies to optimize for multi-user capacity and device density, this level of generality and information extraction to make sense out of all wireless activity is critically important for urban spectrum optimization, coordination with neighboring wireless devices, interferer identification, and identification of security threats and malicious device and activity identification.

\fortim{AD in Wireless networks is a key problem in understanding activity related to spectrum access, efficiency, interference, and security.  Unlike wired networks, it involves making sense out of high-rate raw sensor data rather than structured semantic language or protocols, and has historically been cumbersome and lacked generality.  Data-driven edge sensing algorithms such as deep learning-driven object detection schemes have helped to make this feasible at scale, and have enabled large-scale AD and change detection across a wide bandwidth and numerous sensors in a general sense.  
While early approaches to this problem have looked at unsupervised generative models as well as traditional baselines for out-of-distribution detection, these approaches can be challenging in noisy observations, especially with subtle changes in spectrum activity.  In this work, we introduce an approach to leverage contrastive learning, uncertainty quantification, and expert feedback to enhance these generative models in a powerful training framework.  Several unique wireless AD datasets are evaluated and performance is compared in terms of AUROC, AUPRC, and F1 scores.}}

%% file: sections/related_work.tex
Many ML approaches have been developed for anomaly detection across diverse applications. The recent resurgence of deep learning techniques demonstrating their effectiveness across a wide variety of domains has lead to the development of many novel and powerful modeling paradigms like generative adversarial networks (GAN)~\cite{goodfellow2014generative}, self-supervised representation learning~\cite{jing2020self} and contrastive learning (CL)~\cite{chopra2005learning}. 
\par\noindent
\emph{Contrastive Learning (CL)} imposes structure on the latent space by encouraging similarity in representations learned for \emph{related} instances and dissimilarity in representations for unrelated instances. Such techniques have proven effective, especially when combined with self-supervised learning~\cite{jeong2021training,chen2020simple} and also with labeled data~\cite{khosla2020supervised}. CL has demonstrated promising results in image recognition tasks. However, most of these efforts focus on improving representation learning performance on traditional classification tasks and do not specifically focus on AD. \emph{Generative Adversarial Networks (GANs)}~\cite{goodfellow2014generative} are a powerful generative learning paradigm grounded in an adversarial training setup. However, they are fraught with training instability. Recently, improvements have been proposed to stabilize the GAN training setup by employing \emph{Wasserstien} distance functions~\cite{arjovsky2017wasserstein} and gradient penalties on the learned weights.
\par\noindent
\textbf{Deep Learning for Anomaly Detection}: The aforementioned developments in deep learning have led to techniques such as autoencoders and GANs being employed for the ubiquitous and challenging problem of AD. Specifically, in~\cite{zhou2017anomaly}, a deep robust autoencoder (\textit{Robust AE}) model is proposed, inspired by the Robust Principal Component Analysis  technique, for AD with noisy training data. However, this methodology by design requires knowledge of a subset of anomalies during model training and may be considered semi-supervised, and is not directly related to our context of unsupervised AD. Recently, another line of AD research~\cite{schlegl2017unsupervised} proposes employing DCGAN~\cite{goodfellow2014generative} for unsupervised AD. The authors then build upon their previous work to propose \textit{fAnoGAN}~\cite{Schlegl2019}, a two-step encoder-decoder architecture based on DCGANs where the encoder (trained separately) learns to invert the mapping learned by the Generator (i.e., decoder) of the DCGAN model. We employ \textit{fAnoGAN} as one of the baselines for empirical comparison. \\
\textbf{Contrastive Learning for Anomaly Detection}: There are multiple reports of contrastive learning being utilized for AD. \textit{Masked Contrastive Learning} \cite{cho2021masked} is a supervised method that varies the weights of different classes in the contrastive loss function to produce good representations that separate each class. Even though this method shows promise, it requires knowledge of anomaly labels. \textit{Contrasting Shifted Instances} (CSI)~\cite{tack2020csi} and \textit{Mean Shifted Contrastive Loss} \cite{reiss2021mean} are two unsupervised AD methods based on CL. CSI investigates the power of self-supervised CL for detecting out-of-distribution (OOD) data by using distributionally shifted variations of input data. We employ CSI as one of our baselines. \textit{Mean Shifted Contrastive Loss} applies a contrastive loss modified using the mean representation on representations generated using models pre-trained on ImageNet data. However, this model is not useful for wireless AD as it is pre-trained on a particular kind of data. Also, none of these methods provide a means to incorporate expert feedback.\\
\textbf{Incorporating Expert Feedback}: The solutions presented in \cite{das2016incorporating,gornitz2013toward,pang2018learning,pang2019deep,pang2021toward} all employ human feedback in various ways. Active Anomaly Discovery (AAD) \cite{das2016incorporating} is designed to operate in an anomaly exploration loop where the algorithm selects data to be presented to experts and also provides a means to incorporate feedback into the model. However, its performance is dependent on the number of feedback loops that can be afforded. Hence, such a method could not be applied to wireless AD where the volume of input data is really high. RAMODO \cite{pang2018learning}, combines representation learning and outlier detection in a single objective function. It utilizes pseudo labels generated by other state-of-the-art outlier detection methods and Chebyshev’s inequality. This dependence on other methods to generate pseudo labels can sometimes be unreliable in cases where state-of-the-art outlier detection methods perform poorly. SAAD \cite{gornitz2013toward}, DevNet \cite{pang2019deep} and DPLAN \cite{pang2021toward} are semi-supervised methods, all of which require minimal labeled anomalies and are not suitable for our problem.\\
The advantage of using contrastive learning for AD is that it can be utilized in a self-supervised setup. That is, we can augment the training samples to generate anomalous samples that are very close to the training distribution and utilize them as negative samples in contrastive loss. This allows our model to detect unseen anomalies effectively. Also, the penultimate layer of the GAN discriminators has recently been shown to act as good representations of the input data \cite{Schlegl2019,han2020gan,choi2020gan,cheng2020adgan}. Hence, the combination of these powerful techniques, CL and GAN serve well for our AD task. None of the related approaches outlined above have developed AD techniques that combine the aforementioned techniques for AD. Also, none of the state-of-the-art related AD approaches provide a means to incorporate expert feedback via contrastive learning.

%% file: sections/background.tex
We propose \ourmethod and \ourframework which employ techniques such as adversarial learning, contrastive learning (CL) and uncertainty quantification (UQ). We shall now briefly introduce these concepts before detailing the full~\ourframework framework in section~\ref{sec:problem_formulation}.

\subsection{Generative Adversarial Networks (GAN)} 
GANs are a class of generative models where the learning problem is formulated as a game between two neural networks, namely the \emph{generator} (G) and the \emph{discriminator} (D). The problem setup of GANs comprises the generator learning to transform inputs sampled from a noise distribution into a distribution $P_f$ such that it resembles the true data distribution $P_r$. Essentially, the generator is trained to \emph{fool} the discriminator while the discriminator is tasked with distinguishing between \emph{fake} samples $\tilde{x} \thicksim P_f$ generated by G and \emph{real} samples $x \thicksim P_r$. 
\hide{GAN \cite{goodfellow2014generative} is an adversarial framework that consists of two networks, a generator, and a discriminator. The task of the generator during training is to reproduce the input in a realistic manner and the task of the discriminator is to identify if the input to the discriminator is real or the one constructed by the generator. Let us consider input which has a  distribution $P_{real}$ and output of the generator has a distribution $P_{fake}$.} The traditional GAN~\cite{goodfellow2014generative} setup minimizes the Jenson-Shannon (JS) divergence between $P_{r}$ and $P_{f}$. However, this divergence is not continuous with respect to the parameters of G, leading to training instabilities. Wasserstein GAN \cite{arjovsky2017wasserstein} (WGAN) was proposed to address this issue. WGAN employs Earth-Mover distance (instead of the JS divergence) which under mild assumptions does not have discontinuities and is almost universally differentiable. Consider the discriminator (also termed the \emph{critic}\footnote{words `critic', `discriminator' are used interchangeably in the paper.} in WGAN) D parameterized by $\Omega$ and generator G parameterized by $\theta$. Eq.~\ref{eq:wgan} depicts the WGAN loss function where  $D$ is parameterized by $\Omega \in \mathcal{B}$, where $\mathcal{B}$ is the set of 1-Lipschitz functions. \hide{where family of functions with $\|D_{\Omega}\|_{L} \leqslant K$ which means $D_\Omega$ should have a Lipschitz constant less than K. This constraint was enforced during training with weight-clipping of weights of the discriminator. However, while weight-clipping is enforced it was found that the critic learned simple functions and did not account for variation in data.} 

\begin{equation}
    \small
    L^{w} = \min_{\theta}\max_{\Omega \in \mathcal{B}} \underset{{x \sim P_r}}{\mathbb{E}} [D_\Omega(x)] - \underset{{\tilde{x} \sim P_f}}{\mathbb{E}} [D_\Omega(\tilde{x})]
    \label{eq:wgan}
\end{equation}

 Enforcing the 1-Lipschitz constraint on $D_{\Omega}$ has been found to be challenging. On the basis of the property that a function is 1-Lipschitz if and only if it has a norm no greater than 1 everywhere,~\cite{gulrajani2017improved} proposed a solution of augmenting the WGAN loss with a soft constraint enforcing that the norm of the discriminator gradients (w.r.t the inputs) be 1. 
 The objective function of the WGAN with this updated soft constraint (termed a gradient penalty) is shown in Eq.~\ref{eq:wgangp}.
\begin{equation}
\small
 L^{gp} = L^{w} + \lambda \; \displaystyle \mathop{\mathbb{E}}_{\check{x} \sim P_{i}} \left( \|\nabla D_\Omega(\check{x})\|_2 -1 \right)^2  
 \label{eq:wgangp}
\end{equation}
Each sample $\check{x} \sim P_{i}$ is generated as a convex combination of points from $P_r$, $P_f$ (i.e., sampled from the line connecting points from $P_r$, $P_f$). $\lambda$ enforces the strictness of the gradient penalty.

\subsection{Contrastive Learning (CL)} \label{supcon}
The paradigm of contrastive learning (CL) has recently demonstrated highly effective results across a diverse set of disciplines and tasks, especially in computer vision~\cite{chen2020simple,chopra2005learning,zbontar2021barlow}. The goal of CL is to impose structure on latent representations learned by a model (M). This is often achieved using soft penalties (e.g., additional loss terms) that influence representations generated by M to be structured so representations of \emph{related} instances are closer together relative to instances that are known to be \emph{unrelated}.\hide{CL losses are primarily imposed to govern pair-wise representation proximity between instances.} Most CL losses are set in a \emph{self-supervised} context where \emph{relatedness} is generated via augmentations of an instance and two distinct instances are considered to be unrelated. 
\hide{SimCLR \cite{chen2020simple} is a self-supervised contrastive learning framework that uses contrastive loss to bring representations of the augmentations of the same instances close to one another. By doing this, they explain that the learned representations from the self-supervised task serve better in downstream tasks such as classification. Supervised contrastive learning \cite{khosla2020supervised} (SupCon) is a variant of SimCLR which specifically addresses the supervised scenario using a different format of the contrastive loss proposed by SimCLR.} 

Recently,~\cite{khosla2020supervised} proposed \emph{supervised contrastive learning} (SupCon), which is an extension of the CL paradigm to supervised (classification) settings. A model trained with SupCon on a labeled dataset learns latent representations grouped by \emph{class labels} while also forcing separation in representations between instances belonging to different classes (i.e., low intra-class separation and high inter-class separation of latent representations).
\par\noindent
Consider a dataset of instances 
$\mathcal{D} = \{(x_1,y_1),..,(x_m,y_m)\}$ such that $x_i \in \mathbb{R}^{b\times l}$ and $y_i \in \mathcal{C}$ is the label of $x_i$ and $\mathcal{C}$ is the set of class labels. Then, the supervised contrastive loss is defined by Eq.~\ref{eq:supcon}.
\begin{equation}
\small
L^{sup} =    \sum_{x_i\in{\mathcal{D}}} \frac{-1}{|Pos(i)|} \sum_{x_{k}\in{Pos(i)}} \log \frac{\exp \left(z_i \cdot z_{k} / \tau \right)}{\sum_{j\in{Q(i)}}{\exp \left(z_i \cdot z_j / \tau \right)}}     \\
\label{eq:supcon}
\end{equation}
Here, $z_i \in \mathbb{R}^{h\times 1}$ is the latent representation of $x_i$ generated by model M. $\textrm{Pos}(i) = \{x_k \in \mathcal{D} | y_k == y_i \land k\neq i\}$ is the set of instances that form the `positive set' for $x_i$. $Q(i) = \{D\backslash x_i\}$.
$\tau \in \mathbb{R}^+$ is a hyperparameter. We employ Eq.~\ref{eq:supcon} for CL but with labels generated in a self-supervised manner.

\hide{Where $z_i=P_r(x_i)$ is a representation of $x_i$, $P_r(x_i) \subset S \setminus x_i $ represents samples that have the same class as $i$, $Q \equiv S \setminus i$,  and $\cdot$ indicates the inner dot product.} 

\hide{The above loss function calculates contrastive loss for $\forall \: x_i \in S$. In a similar manner, we can also calculate the contrastive loss for a particular class $c$ using the following. \\

\begin{equation}
\small
L^{supclass}(S,c) =  L^{sup}(S) \; \textrm{where} \; \; \forall x_i : y_{x_i} = c    \\
\end{equation}
}

\subsection{Uncertainty Quantification (UQ)}\label{sec:uq_background}

Quantifying decision uncertainty is critical to the success of real-world machine learning (ML) frameworks. It is of special relevance in the current setting of anomaly detection wherein the confidence of a model in its decision additionally indicates the urgency of a potential alert issued by the model. While traditional ML models yield point predictions, Bayesian ML provides a framework for capturing model uncertainty. One such UQ approach~\cite{gal2016dropout}, can be considered to approximate Gaussian Processes with neural network models. This approach termed Monte-Carlo Dropout entails running a monte-carlo sampling (during inference) of a trained model by randomly masking a set of learned weights of the model each time (i.e., \emph{dropout}~\cite{srivastava2014dropout}). This is akin to sampling from the \emph{approximate posterior} which leads to uncovering the model's predictive distribution. Inferring the predictive distribution is one of the methods to quantify model uncertainty with Bayesian neural networks.

\hide{In anomaly detection, when a model calls something an anomaly or not, that is, a binary 0 or 1, when the model transitions from desk to a real-world setup, there could be a lot of false positives or false negatives. However, it is very important to not have such misclassifications in mission-critical tasks. This brings the need to quantify and present the amount of uncertainty of a prediction to the end user. Uncertainty Quantification is an important component for transitioning from research to application. Monte Carlo Dropout (MC Dropout) \cite{gal2016dropout} is a method to quantify uncertainty. Dropout \cite{srivastava2014dropout} is a regularization technique that removes connections in a deep learning model based on probability. This removal helps the model rely less on individual nodes and helps the weights to be learned together. In a regular dropout scenario, dropout is used only during training and dropped during inference. However, in MC dropout, the dropout layer is kept active during inference. This introduces a degree of uncertainty in predictions that MC dropout exploits. For example, if we run one test instance through the model 10 times, we will get 10 different predictions and the proportion of correct predictions can be used as an uncertainty measure.We utilize this method in \ourmethodUQ and \ourframework.}

%% file: sections/problem_formulation.tex

\begin{figure*}[!ht]
    \centering
    \includegraphics[width=0.7\textwidth]{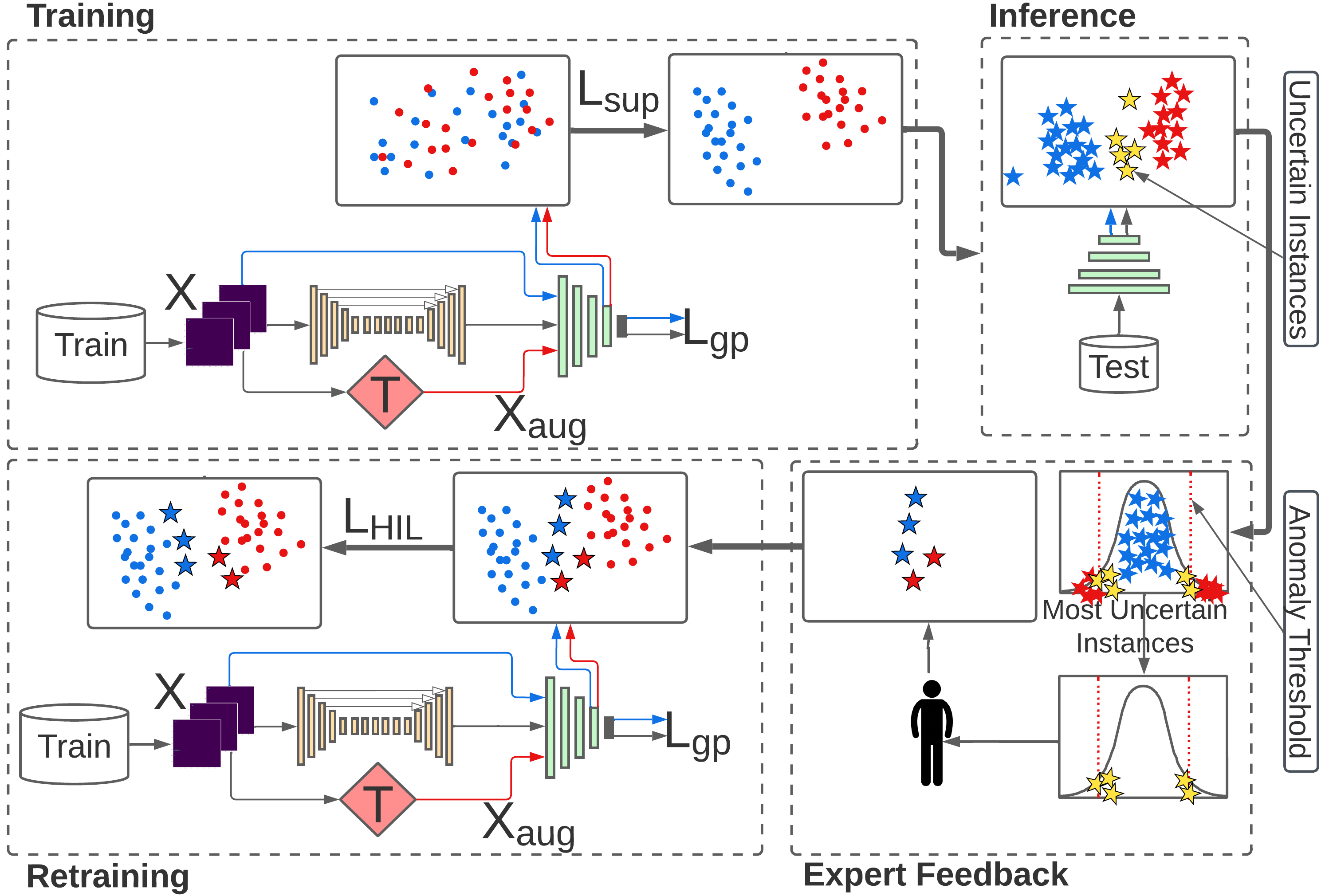}
    \caption{The full architecture of the human-in-the-loop \ourframework anomaly detection framework. (Training): The framework consists of a WGAN-GP with an uncertainty-aware discriminator trained with supervised contrastive learning (SupCon) to impose structure in the latent space. Labeled data required for SupCon is obtained by applying `negative transformations' on a benign set of instances to generate corresponding anomalous instances. (Inference): During inference, the model yields a prediction (\emph{anomaly}:red star or \emph{benign}:blue star) for every instance, accompanied by the prediction uncertainty. (Expert Feedback): Uncertain instances (yellow stars) are isolated and passed to an \emph{expert} to uncover their true labels. (Re-Training): The pre-trained WGAN-GP model is then fine-tuned with this additional expert feedback to further improve its representations learned thereby leading to improved anomaly detection performance and decreased prediction uncertainty.}
    \label{fig:architecture}
\end{figure*}

\hide{Our objective is to create a model that can learn good representations which can be used to distinguish between benign and anomalous instances in a network wireless anomaly detection setup. The penultimate layer of the discriminator in GAN is known to learn representations that are reliable for detecting anomalous instances \cite{Schlegl2019}. Therefore we formulate a method to enhance the representations in the penultimate layer in the discriminator that can help us achieve our objective.} 
We shall now describe the various facets of our novel human-in-the-loop anomaly detection framework~\ourframework. Fig.~\ref{fig:architecture} details the overall architecture of~\ourframework.

\subsection{Self Supervised Anomaly Detection with Negative Transformations}\label{sec:caad_training}
The core of the proposed framework is the Contrastive Adversarial Anomaly Detection (\ourmethod) model. The structure of the~\ourmethod model resembles a WGAN with gradient penalty (WGAN-GP) as described in section~\ref{sec:background} comprising a generator G$_\theta$ and a discriminator D$_\Omega$. In addition to GAN-based training, we also train~\ourmethod discriminator with CL to impose explicit structure on the learned latent representations and improve representation learning. The CL technique employed is similar to supervised contrastive learning (SupCon) detailed in section~\ref{sec:background}. However, SupCon requires a labeled dataset. To generate a labeled dataset $\mathcal{D}$, we assume the existence of a training set without any anomalies. Let this set be denoted $\mathcal{D}_b = \{(x_1,y_1),..,(x_m,y_m)\}$, such that $y_i = 0,\, \forall (x_i,y_i) \in \mathcal{D}_b$. We apply a negative transformation $T(\cdot)$ to violate the normalcy of every instance $x_i \in \mathcal{D}_b$ to obtain a corresponding set of anomalous instances $\mathcal{D}_a = \{(x_1,y_1),..,(x_m,y_m)\}$ such that $y_i = 1,\, \forall (x_i,y_i) \in \mathcal{D}_a$. Now let us consider $\mathcal{D} = \{\mathcal{D}_a,\mathcal{D}_b\} = \{(x^a_1,y^a_1),..,(x^a_m,y^a_m),(x^b_1,y^b_1),..,(x^b_m,y^b_m) | y_i^a = 0 \land y_i^b = 1 \, \forall i\}$. Then we can leverage Eq.~\ref{eq:supcon} to directly train the~\ourmethod discriminator with CL (specifically SupCon).
\begin{equation}
    L^{CAAD} = L^{gp} + \alpha L^{sup}
    \label{eq:caad_loss}
\end{equation}
Eq.~\ref{eq:caad_loss} represents the objective employed to train the~\ourmethod  model to learn effective representations of benign and negatively transformed `anomalous' instances via CL. Here, $\alpha$ governs the effect of the supervised contrastive loss on the discriminator representations.
\hide{Consider $P_r$ which represents part of the discriminator up to the penultimate layer. $d_i = P_r(x_i)$ where $x_i$ is the input instance, gives us the 800 dimensional embeddings we are interested in. For a given set of $X=[x_i]^{N}_{i=1}$ instances, we apply a set of negative transformations $T$ and obtain $X^T = T(X)$. Consequently, the contrastive loss is formulated in Eq \ref{eq:caadcontloss}. }

\hide{\begin{equation}
    \small
    L^{cont} = L^{sup}(\{X,X^T\})  
        \label{eq:caadcontloss}
\end{equation}
\begin{equation}
    \small
    \textrm{where } Pos(x_i) = \begin{cases} X \setminus x_i &\text{if }x_i \in X \\ X^T \setminus x_i &\text{if }x_i \in X^T \end{cases}
\end{equation}

Now, we have formulated the contrastive loss and the GAN loss. Hence, the overall loss function for \ourmethod is given in Eq. \ref{eq:caadloss}.
\begin{equation}
    \small
    L_{D} =  L^{gp} + \alpha L^{cont}
    \label{eq:caadloss}
\end{equation}
where $\alpha$ is the weighting coefficient.}
\subsection{Inferring Decision Uncertainty with~\ourmethodUQ}
In order to maximize the effect of expert feedback on model performance, we are required to isolate an \emph{effective} set of instances for which to solicit feedback. To this end, we define this \emph{effective} set of instances as those for which the model is the most uncertain in its prediction. Hence ~\ourmethod, trained to learn effective representations as defined in section~\ref{sec:caad_training}, is augmented to quantify its prediction uncertainty using the popular Monte-Carlo dropout technique (see section~\ref{sec:uq_background}). This variant of~\ourmethod augmented with uncertainty quantification capability is termed~\ourmethodUQ. Concretely, the model structure of~\ourmethod discriminator is augmented by including \emph{dropout} in each layer of the model to yield~\ourmethodUQ. Let $D^L_{\Omega^*}$ represent the first $L$ layers of the discriminator of a trained~\ourmethodUQ model where training happens according to Eq.~\ref{eq:caad_loss}. Further, let $d_i = D^L_{\Omega^*}(x_i)$, then, $\{d^j_i\}_{j=1...k}$ represents the set of `k' monte-carlo sampled embeddings obtained from $D^L_{\Omega^*}(x_i)$.
\hide{Now we have a model that can learn good representations for identifying anomalies. We further extend \ourmethod to produce an uncertainty measure for each prediction. Since this is \ourmethod with uncertainty quantification, we call this model \ourmethodUQ. We measure uncertainty by utilizing the MC dropout. We add dropout layers after each of the convolution layers in the discriminator and train the model. Let $d_{i} = P_r(x_i)$. $\{d^j_i\}_{j=1..k}$ is a set of discriminator embeddings produced by $k$ monte carlo inferences on instance $x_i$.} \ourmethodUQ employs the mean of the Monte-Carlo embeddings, denoted $\overline{d}_i$ as the representation inferred for an instance $x_i \in \mathcal{D}$.

Every Monte-Carlo embedding $d^j_i$ generated by $D^L_{\Omega^*}(x_i)$ is subjected to a \emph{scoring mechanism} (section~\ref{sec:anom_scoring}) whereby a prediction $\hat{y}^j_i \in \{0,1\}$ is obtained. Here $\hat{y}^j_i = 0$ indicates a benign classification and $\hat{y}^j_i = 1$ indicates an anomalous classification of $x_i$ at MC sample $j$. Prediction uncertainty as quantified by~\ourmethodUQ for $\overline{d}_i$ is outlined in Eq.~\ref{eq:pred_uncertainty}

\begin{equation}
    u_{i,c} = |\{\hat{y}_i^j |j \in \{1,2,..k\} \land \hat{y}_i^j = c\}| \; \text{where } c\in\{0,1\}\\
\label{eq:pred_uncertainty_pre}
\end{equation}
\begin{equation}
    \mu_i = 1 - \frac{max({u_{i,0},u_{i,1}})}{k}
\label{eq:pred_uncertainty}
\end{equation}

\hide{Hence, for $k$ inferences on $x_i$, we have $\{\hat{y}^j_i\} \equiv{\{\{\hat y_{i,j,ben}\},\{\hat y_{i,j,anom}\}\}}$ where $\{y_{i,j,anom}\}$ are the anomaly predictions and $\{y_{i,j,ben}\}$ are the benign predictions. We calculate uncertainty as
\begin{equation}
\small
    \mu_{x_i} = \frac{|\{y_{i,j,anom}\}|}{k}
\end{equation}

\begin{equation}
    \mu_{i} = \frac{|\{\hat{y}_i^j |j \in \{1,2,..k\} \land \hat{y}_i^j = 1\}|}{k}
    \label{eq:pred_uncertainty}
\end{equation}
In Eq.~\ref{eq:pred_uncertainty}, $|\cdot|$ indicates cardinality and $\mu_i\in{[0,1]}$ is the uncertainty measure. $\mu_i \approx 0$ indicates a high confidence benign classification and $\mu_i \approx 1$ indicates a high confidence anomaly classification of instance $x_i$ by our model.
that the prediction is very certain that an instance is benign and 1 is very certain that an instance is an anomaly.}

\subsection{Leveraging Expert Feedback}

Now, we have formulated a method to calculate an uncertainty measure $\mu_i$ for any instance $x_i$. For $\{x_i\}\in\mathcal{D}$ for which we want to make predictions, if $\mu_i \thickapprox 1 \; \forall \; x_i$, then we can be sure that the model predictions are reliable. However, if this is not true, we further retrain \ourmethodUQ for a small number of epochs using a set of \emph{effective} instances determined using $\mu_i$ and their corresponding feedback from an expert on these instances, along with the original training set. We call this model \ourframework since this is a contrastive adversarial anomaly detection model trained from expert feedback. 

For a particular class $c$, we define the supervised contrastive loss in Eq. \ref{eq:supconclass}. 
\begin{equation}
    L^{supclass}(\mathcal{D},c) =  L^{sup}(\mathcal{D})  \; \; \forall x_i : y_{i} = c    \\
\label{eq:supconclass}
\end{equation}

$L^{supclass}$ is used to only bring instances of one class $c$ together and away from all other classes, in contrast with $L^{Sup}$ which brings each instance close to each other instance of the same class and away from instances of all other classes. 

Let X denote a set of benign instances. From the set of inferences yielded by \ourmethodUQ, we select the top `h'\% most uncertain instances $X^{HIL}$, based on $\mu_i$ (Eq. \ref{eq:pred_uncertainty}) as the \emph{effective} set of instances and showcase them to an expert to receive feedback. This feedback gives us $\{X^{HIL}_{anom}, X^{HIL}_{ben}\}$ were $X^{HIL}_{anom}$ and $X^{HIL}_{ben}$ are the set of instances labeled by an expert as anomalous and benign respectively. We then incorporate an additional loss term in the loss function of \ourmethodUQ and retrain the model for a small number of epochs. Let $X_{aug}=T(X)$ where $T$ is a class of transformations, $\mathcal{D}_1 = \{X^{HIL}_{anom},X\}$, $\mathcal{D}_2 = \{X^{HIL}_{ben},X_{aug}\}$, $\mathcal{D}_3 = \{X^{HIL}_{ben},X^{HIL}_{anom}\}$. We define the HIL loss $L^{HIL}$ in Eq \ref{eq:hilloss}.  


\begin{equation} \label{eq:hilloss}
\begin{split}
L^{HIL} & = \alpha_1 L^{supclass}(\mathcal{D}_1,c=1)
  + \alpha_2 L^{supclass}(\mathcal{D}_2,c=0) \\
 & + \alpha_3 L^{supclass}(\mathcal{D}_3,c=0)
\end{split}
\end{equation}

where the first term in $L^{HIL}$ helps bring $X^{HIL}_{anom}$ together while also pushing it far away from $X$, the second term helps bring $X^{HIL}_{ben}$ together while pushing it far away from $X_{aug}$ and the third term helps bring $X^{HIL}_{ben}$ together and pushes it away from $X^{HIL}_{anom}$. Hence the overall loss term for the retraining model \ourframework is given below.
\begin{equation}
    L_{D} = L^{CAAD} + L^{HIL}
\end{equation}

\subsection{Anomaly Detection}\label{sec:anom_scoring}
Training our model gives us meaningful embeddings. Here we define how we use these embeddings for identifying anomalies.\\
\textbf{Scoring function:} We define a scoring function that can be used to determine if an instance is an anomaly or not. We adopt the scoring mechanism used in~\cite{tack2020csi}. Consider a set of instances used during training. We cluster them into $m$ different clusters and obtain their cluster centroids as $\{x_m\}$. For every test instance $x_i$, the score is calculated as below.
\begin{equation}
    s_{x_i} = max(cosine(D^L_{\Omega^*}(x_i),D^L_{\Omega^*}(x_m)))  \; \forall \; x_m
\end{equation}

\textbf{Anomaly threshold:} 
Consider a validation set ${x_v}$ and a distribution P of anomaly scores $s_{x_v}$.
\begin{equation}
    \theta = \arg_\theta \{P(s_{x_v}<\theta)= \phi \}
\end{equation}
where $\phi$ is the strictness parameter which can be tuned to control the rate of false positives and false negatives. When $s_{x_i}$ exceeds $\theta$, then we call $x_i$ an anomaly.

%% file: sections/experimental_setup.tex

We consider several wireless emission activity datasets as well as the well-known MNIST computer vision dataset for evaluation.  The wireless emission activity datasets consist of metadata describing detected radio emissions observed over the air in a known radio frequency (RF) environment.  Metadata describes a range of aspects including detection time, frequency, bandwidth, signal type, and signal power, and is streamed from an edge sensor into an Elasticsearch database for archival.  Anomalies consist primarily of new emitters coming online or exhibiting new behavior (e.g. hopping) in a band with otherwise orderly patterned behavior, or the disappearance (e.g. failure) of emitters that are otherwise regularly present.

\subsection{Dataset Description}\label{sec:dataset_description}
\par\noindent
\textbf{\datakop}: Long-term wireless emission metadata in the 800-900MHz band collected for a span of 2 months from a first geographical location with a frequency scanning receiver.  We consider a 2D 80x80 bin density feature formed from quantized bandwidth and frequency features from these emissions over 3-minute intervals.  This forms 11670,~5002,~7146 time intervals in a sequential manner for training, validation, and test sets respectively. We denoise training and validation set densities by setting values of bins with a probability of occurrence less than 0.0005 to zero. \hide{For the test set, we consider these low-density noisy observations with probability \< 0.0005 as anomalies, and we additionally inject synthetic anomalies (i.e. additional out-of-distribution emission metadata).} The test set contains 3894 benign and 3738 anomalies which also includes hopping anomalies.
\par\noindent
\textbf{\datamltest}: Long-term wireless emission metadata (similar to \datakop) collected from a second geographical location. \hide{The same methodology for masking low-density noise and injecting synthetic anomalies is used.} Input is preprocessed similar to 80x80 bin density features of \datakop. Data spans a time interval of 10 days and forms 1645,~706,~1310 sequential time intervals for training, validation, and test sets respectively..\hide{In this dataset we do no do denoising, as this dataset was collected especially for the purpose of testing.} The test set contains 786 benign and 724 anomalies.
\par\noindent
\textbf{\dataeighteighty}: Short-term wireless emission metadata collected from a third geographical location comprising short-time high-rate observations of 900MHz cellular and ISM bands. Input similar to 80x80 bin density features of \datakop but with a 1-second interval. One 4G LTE signal (i.e. cellular base station) goes offline (signal drop) at around 198 seconds into the dataset. 154, ~39, ~198 are the number of training, validation, and testing instances respectively. The test set contains 94 benign and 104 anomalies.\\
We note that the datasets \datakop,\, \datamltest,\, \dataeighteighty are collected using on-the-ground sensors and depict real-world benign and anomalous behavior of communication network traffic.
\par\noindent
\textbf{MNIST}: This is a standard image dataset~\cite{deng2012mnist} for machine learning research comprising images of hand-written digits. The training and validation set consists of 4089 and 1753 images of the number `4' (benign class) respectively. During testing, all other classes of digits are considered anomalies. The test set contains 982 benign and 9018 anomalous images leading to an imbalanced evaluation setup.

\subsection{Baselines}
We evaluated several baselines that are either state-of-the-art or closely related to \ourmethod. All of them are explained in this section.
\label{sec:baselines}
\par\noindent
\textbf{Isolation Forest} \cite{liu2008isolation}: A popular benchmark ensemble method for anomaly detection wherein partitions are created in data such that each data point is isolated. During such partitioning, an anomalous data instance isolates itself much easier compared to a benign instance. We employed 100 base estimators.
\par\noindent
\textbf{One Class SVM (OC-SVM)} \cite{wang2004anomaly}:\hide{Support vector machines (SVM) learn a hyperplane which separates two classes of points.} OC-SVMs learn a hypersphere that encompasses points from a single class and any point that falls outside of this hypersphere is considered an outlier. The kernel that was used in the algorithms is a radial basis function kernel.
\par\noindent
\textbf{UNetGAN} \cite{ronneberger2015unet}: GAN model with UNet as a generator, and convolutional network used for the discriminator. The GAN model is trained using Wasserstein loss and gradient penalty.
\par\noindent
\textbf{fAnoGAN}\cite{Schlegl2019}: A state-of-the-art anomaly detection model based on Wasserstein GAN-GP. It comprises a two-stage AD framework wherein the first stage involves WGAN-GP being trained to translate samples from a noise distribution to outputs of the process of interest. The second stage employs the pre-trained WGAN-GP to train an encoder network to learn an `inverse mapping' of the WGAN-GP generator. The anomaly score is calculated as a function of error between intermediate representations of input data and generator output.
\par\noindent
\textbf{Contrasting Shifted Instances (CSI)}\cite{tack2020csi}: A state-of-the-art anomaly detection deep neural network model which employs traditional self-supervised contrastive learning with a novel contrastive task comparing an instance to \textit{distributionally shifted} versions of itself. Anomaly detection occurs through a novel anomaly scoring mechanism.\\
After finding the anomaly score, all the baselines and \ourmethod follow the same procedure as mentioned in \ref{sec:anom_scoring} to detect anomalies.
\subsection{Evaluation Metrics}\label{sec:eval_metrics}
We employ multiple evaluation metrics to provide a holistic quantitative evaluation of model performance on the task of anomaly detection. Specifically, we separately report the \emph{F1 scores} for correctly detecting benign and anomaly instances as well as a \emph{weighted average} of the two F1 scores. Further, we also report the Area Under the Receiver Operating Characteristic (AUROC) metric which is explicitly dependent on the false positive rates (FPR) of the models. We explicitly report AUROC to investigate whether models have low FPRs (thereby high AUROC) values which are imperative for an effective anomaly detection model. Finally, as we also deal with imbalanced datasets (see section~\ref{sec:dataset_description}), we also monitor the Area Under the Precision-Recall Curve (AUPRC) metric which is known to complement AUROC well by alleviating biases due to data imbalance. 

\subsection{Model \& Training Details}

\textbf{Model details:} The backbone of \ourframework is a WGAN with gradient penalty (WGAN-GP). The generator of this WGAN-GP is a UNet autoencoder comprising 5 down convolutional layers (each with kernel size 4, batch normalization, and leaky-ReLU), 5 same convolutional layers (each with kernel size 3, maxpooling, batch normalization, and leaky-ReLU) and 5 up-convolutional layers (each with kernel size 4 batch normalization and ReLU). \hide{It has 5 down-convolution layers, each followed by batch normalization and leaky-relu activation with a negative slope of 0.2. This is followed by 5 same convolutions with each convolution followed by maxpooling, batch normalization, and leaky relu again with a negative slope of 0.2. Finally, we have 5 up-convolution layers, each followed by batch normalization and relu activation. We use a kernel size of 4 for the up and down convolutions and a kernel size of 3 for the same convolutions.} Discriminator has five convolutional layers, each followed by instance normalization, leaky-ReLU (negative slope of 0.2), and a dropout layer with a dropout probability of 0.5.

\textbf{Training details:}
Our models are trained for 100 epochs with a batch size of 32, Adam optimizer ($\beta1=0, \beta2=0.9$), and learning rate of $1e^{-4}$ for both the generator and the discriminator. For the gradient penalty, we use a $\lambda$ value of 10. The weighting coefficients $\alpha, \alpha_1, \alpha_2 \; \textrm{and} \; \alpha_3$ are set to 1 during training. The negative transformation we have employed is \emph{salt noise} for the network datasets (\datakop,~\datamltest,~\dataeighteighty) and $90^{\circ}$ rotations for MNIST. Strictness parameter $\phi = 0.99$ and number of training clusters $m=1$ are used during validation. 
During retraining, we train the model only for 7 epochs. Considering that feedback from experts is not easy to obtain and is expensive, we use a $h$ value of 5 (corresponding to 5\% of the most uncertain instances) during retraining. Although we only explore single-round expert feedback in our experiments, if more feedback can be afforded, our framework can also be used to support multi-round expert feedback for active learning.

%% file: sections/results_and_discussion.tex


\begin{table*}[ht]
\small
\caption{Summary of Results.}
\begin{center}
\label{tab:summary}
    \begin{tabular}{|c|c|C{1.3cm}|C{1.3cm}|C{1.3cm}|C{1.3cm}|C{1.3cm}|}
        \hline
        \textbf{Data} & \textbf{Model} &  \textbf{Benign F1} & \textbf{Anomaly F1} & \textbf{AUROC} & \textbf{AUPRC} & \textbf{Avg.Wt. F1} \\
       \hline\hline
    \multirow{7}{*}{\datakop}
&	Isolation Forest	&	0.75	&	0.47	&	0.88	&	0.83 & 0.61	\\
\cline{2-7} 
&	OC-SVM	&	0.41 &	0.7	&	0.86	&	0.81 & 0.55	\\
\cline{2-7} 
&	CSI	&	0.75 &	0.2	&	0.61	&	0.53 & 0.48	\\
\cline{2-7} 
&	fAnoGAN	&	0.69	&	0.18	&	0.8	&	0.8	& 0.44 \\
\cline{2-7} 
&	fAnoGAN**	&	0.68	&	0.04	&	0.85	&	0.84 & 0.37	\\
\cline{2-7} 
&	UnetGAN	&	0.74	&	0.41	&	0.86	&	0.89 & 0.58	\\ \cline{2-7}
&	~\ourmethod	&	\textbf{0.93}	&	\textbf{0.9}	&	\textbf{0.97}	&	\textbf{0.97} & \textbf{0.92}	\\

\hline\hline
\multirow{7}{*}{\dataeighteighty}
&	Isolation Forest	&	0.64	&	0	&	0.49	&	0.57 & 0.3	\\
\cline{2-7} 
&	OC-SVM	&	0.59	&	0.79	&	0.97	&	0.98 & 0.7	\\
\cline{2-7} 
&	CSI	&	0.03	&	0.81	&	0.37	&	0.58 & 0.44	\\
\cline{2-7} 
&	fAnoGAN	&	0.85	&	0.72	&	0.95	&	0.93 & 0.78	\\
\cline{2-7} 
&	fAnoGAN**	&	0.83	&	0.79	&	0.96	&	0.63 & 0.81	\\
\cline{2-7} 
&	UnetGAN	&	0.85	&	0.9	 &	\textbf{1.0}	&	\textbf{1.0} & 0.88	\\\cline{2-7}
&	~\ourmethod	&	\textbf{0.92}	&	\textbf{0.94}	&	\textbf{1.0}	&	\textbf{1.0} & \textbf{0.93} \\

\hline\hline
\multirow{7}{*}{\datamltest}
&	Isolation Forest	&	0.75	&	0.02	&	0.63	&	0.71 & 0.46	\\
\cline{2-7} 
&	OC-SVM	&	0.34	&	0.59	&	0.74	&	0.78 & 0.44	\\
\cline{2-7} 
&	CSI	&	\textbf{0.84}	&	0.27	&	0.63	&	0.3 & 0.61 \\
\cline{2-7} 
&	fAnoGAN	&	0.75	&	0.14	&	0.7	&	0.58 & 0.51	\\
\cline{2-7} 
&	fAnoGAN**	&	0.76	&	0.6	&	0.73	&	0.63 & 0.7	\\
\cline{2-7} 
&	UnetGAN	&	0.73	&	0.36	&	0.64	&	0.5	 & 0.58\\\cline{2-7}
&	~\ourmethod	&	0.77	&	\textbf{0.73}	&	\textbf{0.86}	&	\textbf{0.83} & \textbf{0.75}	\\

\hline\hline
\multirow{7}{*}{MNIST}
&	Isolation Forest	&	0.28	&	0.63	&	0.88	&	0.59 & 0.6	\\
\cline{2-7} 
&	OC-SVM	&	0.56	&	0.96	&	0.91	&	0.62 & 0.92	\\
\cline{2-7} 
&	CSI	&	0.55	&	0.9	&	0.9	&	0.81 & 0.87	\\
\cline{2-7} 
&	fAnoGAN	&	0.51	&	0.88	&	\textbf{0.98}	&	\textbf{1.0} & 0.84	\\
\cline{2-7} 
&	fAnoGAN**	&	0.31	&	0.65	&	0.95	&	0.99 & 0.62	\\
\cline{2-7} 
&	UnetGAN	&	0.5	&	0.89	&	0.93	&	0.99 & 0.85	\\\cline{2-7}
&	~\ourmethod	&	\textbf{0.76} &	\textbf{0.97}	& 0.93	&	\textbf{1.0} & \textbf{0.95}\\

\hline
\end{tabular}
\end{center}
\label{tab:multicol}
\end{table*}

We now investigate the performance of our novel~\ourframework anomaly detection framework. Our detailed analysis entails a rigorous quantitative and qualitative performance evaluation.
The specific research questions we ask are as follows:

\par\noindent
$\bullet$ How does our~\ourmethod model perform relative to the existing state-of-the-art (SOTA) for anomaly detection?
\par\noindent
$\bullet$ Can we augment~\ourmethod to successfully incorporate expert feedback (\ourframework) to improve the quality of learned representations?
\par\noindent
$\bullet$ How does each facet of our novel~\ourframework framework contribute towards the overall performance? 

\subsection{~\ourmethod Anomaly Detection Performance}
At the outset, we shall investigate the anomaly detection capability of the~\ourmethod model which forms the backbone of our proposed, novel~\ourframework framework. \hide{The~\ourmethod model \label{sec:problem_formulation} forms the backbone of our proposed~\ourframework framework. Hence, at the outset we shall investigate the anomaly detection capability of the~\ourmethod model.} Specifically, we evaluate the anomaly detection performance of~\ourmethod across four datasets comprising diverse characteristics and anomalies (see section~\ref{sec:dataset_description}).

Table~\ref{tab:summary} details the anomaly detection performance comparison of~\ourmethod with several well-accepted state-of-the-art (SOTA) anomaly detection models.
Across all the datasets and types of anomalies,~\ourmethod achieves a mean performance improvement of \textbf{92.84}\% as evidenced by the anomaly \emph{F1 score} metric.~\ourmethod also achieves an overall mean performance improvement of \textbf{59.39\%} across three of the four datasets where~\ourmethod is the best performing model (i.e., combined performance on benign and anomaly detection) as demonstrated by the weighted average F1 score metric. 
\par\noindent
\textbf{False Positives}: An important facet of a robust and practically useful anomaly detection framework is its ability to minimize `false alarms'. To investigate this behavior, we report AUROC (see~\ref{sec:eval_metrics}) which explicitly is a function of the false positive rate (FPR).

~\ourmethod yields consistently high AUROC values (indicative of its low FPR i.e., it produces very few false alarms).~\ourmethod yields the highest AUROC values in three out of the four datasets. We must note that in the case of the MNIST dataset, the AUROC of~\ourmethod (i.e., \textbf{0.93}) is competitive and amenable for use in real-world AD applications.
\par\noindent
Due to the variegated nature of data imbalance in our experiments (see section~\ref{sec:dataset_description} for data support statistics) we also evaluate the AUPRC metric (as a complement to AUROC under data imbalance). We notice that~\ourmethod is the best performing\footnote{accompanied by fAnoGAN on MNIST, UNetGAN on \dataeighteighty} across all datasets (including MNIST) as per the AUPRC metric.
\hide{Lot of the anomaly detection methods that have been reported until now focus on structure anomalies that are very easy to identify. To detect unstructured anomalies that are subtle in nature with a very stringent threshold is a difficult problem. We prove that our model is able to perform well even in such anomaly scenarios.} 
\par\noindent
\textbf{Network Anomaly Detection:}\hide{We notice that~\ourmethod significantly outperforms all other models in identifying anomalous occurrences in the communication network being monitored. This is specifically evidenced by the anomaly F1 score, AUROC, and AUPRC metrics.}~\ourmethod is able to detect extremely `weak' anomalous signatures associated with activity in irregular parts of the spectrum being monitored. This is specifically evidenced by the superior performance of~\ourmethod on datasets~\datakop and\\ ~\datamltest, both of which contain attack signatures generated by devices that inappropriately access unused regions of the band being monitored. The superior performance of~\ourmethod in anomaly detection on the~\dataeighteighty dataset which consists of the `signal drop' anomaly (described in section~\ref{sec:dataset_description}), also demonstrates the versatility of the~\ourmethod model to detect different types of irregularities in different bands across the communication spectrum. We notice that the CSI model has a higher benign F1 score but lower overall wt.Avg. F1 score (as it underperforms on the corresponding anomaly detection task) for the \datamltest dataset.~\ourmethod in contrast yields more stable results for detecting both benign and anomalous instances across all datasets.  

\par\noindent
\textbf{MNIST Anomaly Detection:} We once again notice that~\ourmethod is able to outperform all other SOTA models for anomaly detection on the MNIST dataset. This result is significant, as it is indicative of the generic nature and flexibility of the proposed solution in addressing variegated anomaly detection tasks. We once again notice that~\ourmethod yields the best anomaly detection F1 score (\textbf{1.04}\% improvement over next best model OC-SVM) and the best weighted average F1 score (\textbf{3.26}\% improvement over next best model OC-SVM).~\ourmethod is also the best performing for the benign instance recognition (indicated by benign F1 score).
The inconsistency in model performance across the benign and AD tasks is once again evident in the context of skewed results in the OC-SVM (fails to detect benign instances accurately). Hence, we once again note (as evidenced by the Wt. Avg. F1 score, Benign F1 score, and Anomaly F1 score) that~\ourmethod yields the best performance on the MNIST AD task.
\hide{kop and mltest contain anomalies that are very subtle and hard to identify. We know this from the number of anomalies present in the huge feature space which is very minimal (10 out of 6400) but also from the poor performance of other models in these datasets which can be seen from the anomaly f1 scores. Specifically in the KOP dataset, \ourmethod has produced a 200\% increase in anomaly F1 score and a 130\% increase in anomaly F1score compared to other baseline methods. Likewise, benign f1 scores, AUROC, and AUPRC values increase by 38\%, 27\%, and 40\% respectively compared to all other baselines. }

\hide{In the network AD task, in the kop dataset \ourmethod achieves an improvement in f1 scores from 0.8 average on all other datasets to 0.97. Only \ourmethod was able to achieve an f1 score of \>0.9 in benign as well as in anomalies.}
\par\noindent
\textbf{SOTA Models}: Standard AD models like Isolation Forest and OC-SVM perform poorly, especially for the challenging \emph{network AD} task. They are unable to identify the subtle anomalous patterns of interest. The fAnoGAN variants showcase unstable performance across the two tasks of benign and AD as evidenced by the large differences in the F1 scores for each dataset thereby rendering them practically ineffective for use as real-world AD frameworks. CSI which is a recent SOTA AD model that also employs contrastive learning significantly underperforms relative to~\ourmethod (avg. performance improvement by Wt. Avg. F1 score \textbf{58.78}\%) across all the datasets. 
\par\noindent
Overall, experiments in Table~\ref{tab:summary} indicate the superior representation learning and AD capability of~\ourmethod on small and large, balanced and imbalanced datasets comprising multiple different types of anomalies.

\hide{Specifically, Isolation forest predicts everything as benign in three of the four datasets, whereas, One-class SVM predicts everything as anomalies in two of the four datasets. 
CSI, which also employs a contrastive loss, predicts all samples as anomalies for 880 and most samples as benign in the kop dataset. The effect of our contrastive loss setup can clearly be seen when the performance of UnetGAN and \ourmethod are compared.



880 dataset has a sample size of only 154. The performance of the Robust autoencoder in this dataset is comparatively worse than other datasets which indicates that it requires more training samples to perform well. \ourmethod completely outperforms Robust Autoencoder in the 880 dataset and 50\% better with respect to benign and anomaly F1 score across all other datasets. 

The effect of our contrastive loss setup can clearly be seen when comparing the performance of UnetGAN and \ourmethod. In all the datasets, \ourmethod produces better anomaly as well as benign f1scores and AUROC and AUPRC scores.

In MNIST, OCSVM has a good anomaly f1score of 0.96. However, the benign f1 score is 0.56 which is poor compared to \ourmethod which has an f1 score of 0.76. \ourmethod also outperforms all the other baselines in the MNIST dataset. This posits that our method does not only apply to network data and anomalies but also works well in the case of structured anomalies.}


\begin{table*}[!h]
    \small
    \centering
    \caption{Impact of fine tuning with expert feedback.}
    \label{tab:anom_detection_results_kop}
    \begin{tabular}{|c|c|C{1.3cm}|C{1.3cm}|C{1.3cm}|C{1.3cm}|C{1.3cm}|}
        \hline
        \textbf{Data} & \textbf{Model} &  \textbf{Benign F1} & \textbf{Anomaly F1} & \textbf{AUROC} & \textbf{AUPRC} & \textbf{Avg.Wt. F1} \\
       \hline\hline
    \multirow{4}{*}{\datakop}
&	\ourmethod	&	0.93	&	0.9	    &	0.97	&	0.97    &	0.92\\
\cline{2-7} 
&	\ourmethodUQ	    &	0.92	&	0.9	    &	0.97	&	0.98    &	0.91\\
\cline{2-7} 
&	\ourframework	    &	0.94	&	0.94	&	0.98	&	0.98    &	0.94\\
\cline{2-7} 
&	\ourframework 95\%	&	0.95	&	0.94	&	0.98	&	0.98    &	0.95\\
            \hline\hline
            \multirow{4}{*}{\dataeighteighty}
&	\ourmethod	&	0.92	&	0.94	&	1	&	1	&   0.93\\
\cline{2-7} 
&	\ourmethodUQ	           	&	0.93	&	0.94	&	1	&	1	&   0.94\\
\cline{2-7} 
&	\ourframework	           	&	0.98	&	0.98	&	1	&	1	&   0.98\\
\cline{2-7} 
&	\ourframework 95\%	       	&	0.98	&	0.99	&	1	&	1	&   0.99\\
        \hline
    \end{tabular}
    \label{tab:expert_feedback}
\end{table*}

\subsection{Anomaly Detection with Expert Feedback} \label{ssec:expertfeedback}
Real-world systems can often benefit from incorporating valuable expert knowledge to influence their representation learning capabilities.
To this end, in an effort to further improve the performance of~\ourmethod, we augment it with the capacity to incorporate expert feedback received in the form of instance labels for a limited set of instances. The resulting framework~\ourframework comprises an augmentation to the discriminator of the~\ourmethod model enabling it to characterize its prediction uncertainties (see section~\ref{sec:problem_formulation}). This uncertainty-aware model (\ourmethodUQ) is trained in a similar fashion to~\ourmethod. Once trained,~\ourmethodUQ yields inferences on unseen instances accompanied by its prediction uncertainty. We select `h\%` of the most uncertain instances as inferred by~\ourmethodUQ to be labeled by an expert. This labeled set of instances is leveraged in a feedback loop to fine-tune the representations learned by~\ourmethodUQ, thus yielding a holistic human-in-the-loop anomaly detection~\ourframework framework.
\begin{figure}[!h]
    \centering
    \includegraphics[width=0.95\columnwidth]{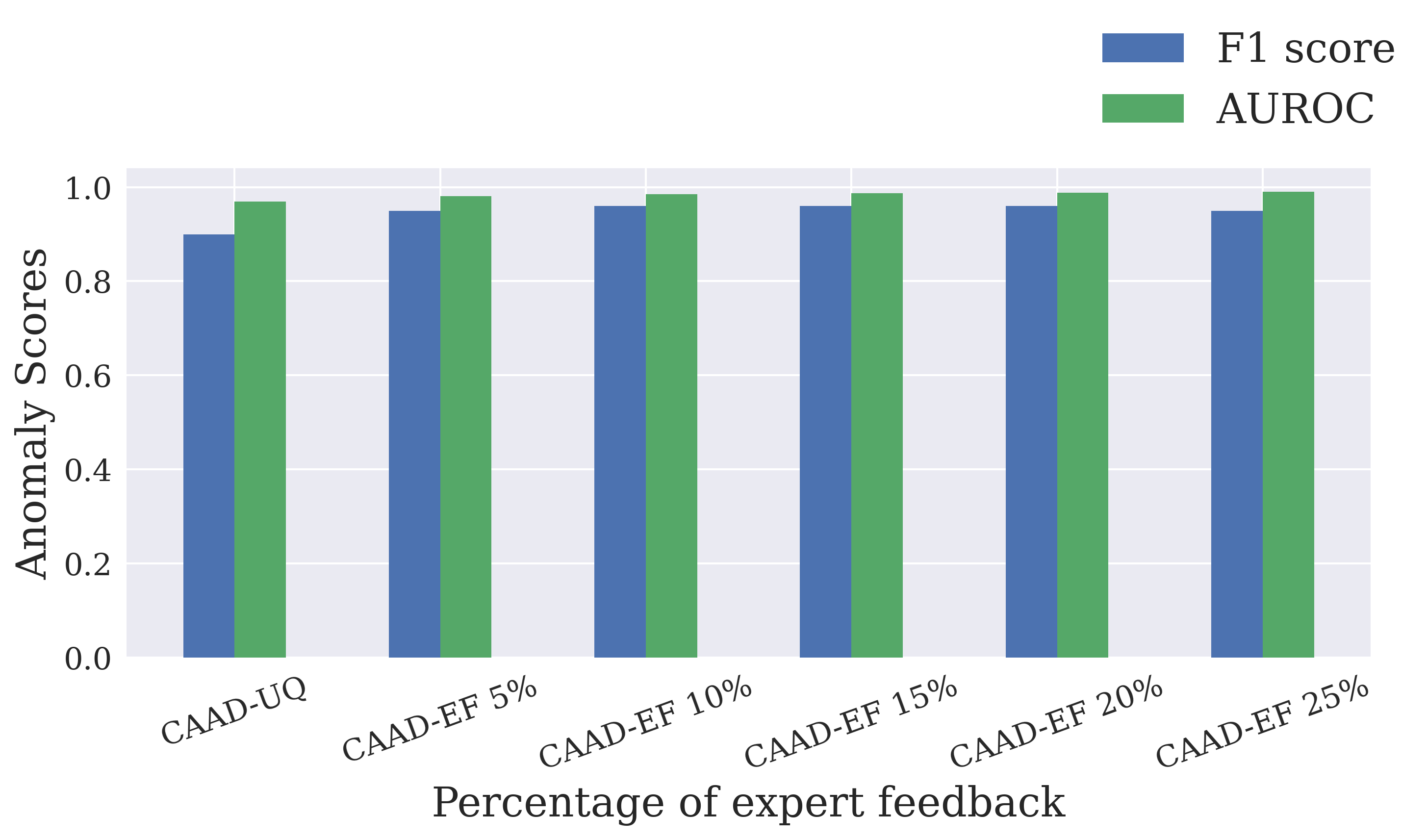}
    \caption{Improvement in Anomaly F1 scores and AUROC values are observed as we increase the percentage of expert feedback 0\% (\ourmethodUQ) to 25\% feedback. We notice that the model is able to show noticeable improvement in the F1 score even with 5\% expert feedback.}
    \label{fig:f1scoreplot}
\end{figure}
\par
\begin{figure}[!h]
    \includegraphics[width=0.99\columnwidth]{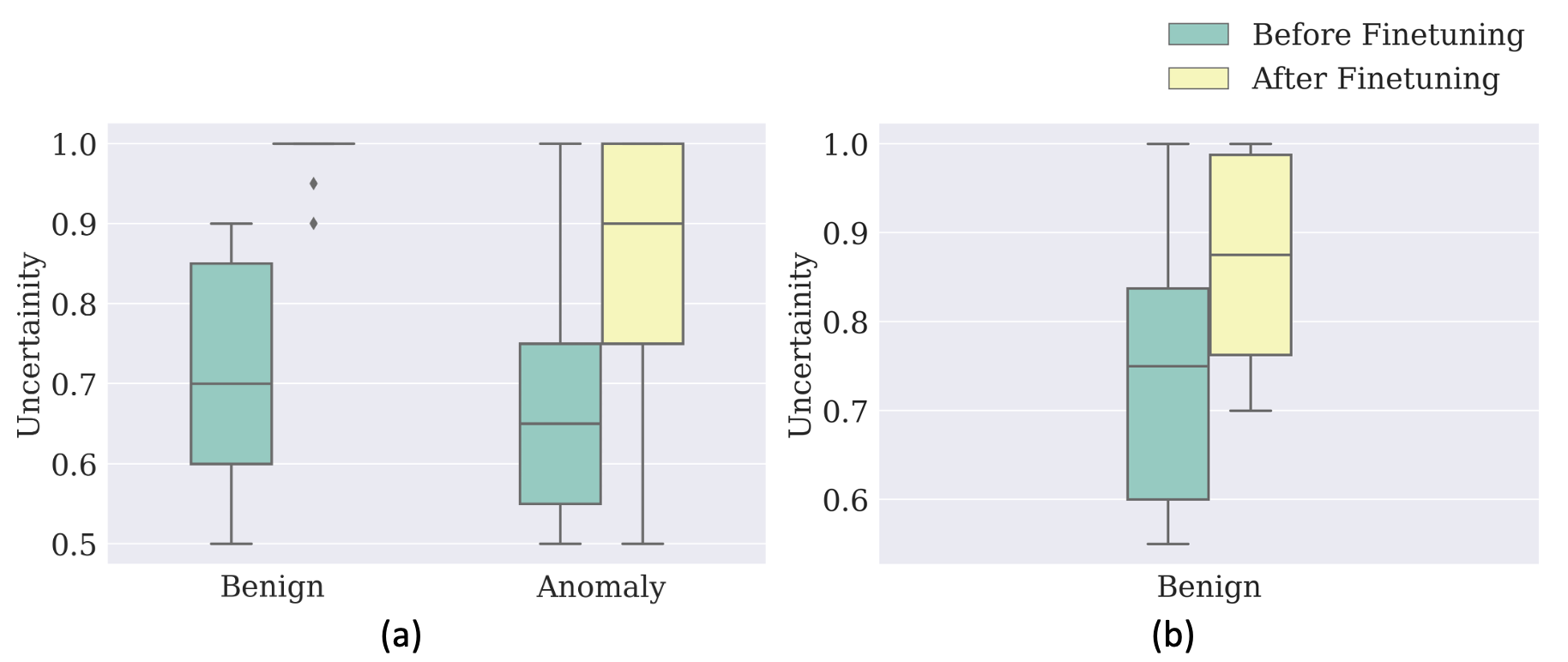}
    \caption{These figures display box plots of uncertainty values of human-in-the-loop (HIL) instances before and after retraining. Figure (a) shows results from the LTW1 dataset and figure (b) shows results from the STW1 dataset. Figure (b) does not contain values for HIL anomalies as there were no HIL anomalies. Here, an uncertainty value of 1 indicates high certainty of prediction and an uncertainty value of 0 indicates low certainty of prediction.}
    \label{fig:uncertainityimprovement}
\end{figure}

\noindent
\textbf{Effect of Expert Feedback (Quantitative Evaluation)}: To investigate the effectiveness of expert feedback in the context of large and small datasets, we select~\datakop and~\dataeighteighty datasets to inspect~\ourframework performance. Table~\ref{tab:expert_feedback} showcases the experimental results. In this table, ~\ourframework 95\% is the ~\ourframework evaluated only on a test set with expert feedback instances removed. We see that incorporating expert feedback (i.e., ~\ourframework) yields a significant performance improvement in the case of large (\datakop) and small (\dataeighteighty) training datasets. Specifically, we notice that incorporating expert feedback on a small subset of uncertain instances (we select instances corresponding to 5\% of the most uncertain test predictions as candidates for expert feedback) yields an average performance improvement of \textbf{4.17}\% over the next best model on the AD task (i.e., improvement in Anomaly F1 score) across both datasets. This indicates that the~\ourframework benefits significantly from expert feedback in the context of different anomalies and data sizes. Additionally,~\ourframework also showcases a \textbf{3.79}\% performance improvement over~\ourmethod (i.e., the variant without explicit feedback) in recognizing benign instances (i.e., improvement in Benign F1 score) across both the datasets showcasing a holistic performance improvement. These results are indicative of a highly effective, holistic and generic AD solution. For completeness, we further characterized the effect of model performance of~\ourframework with the increase in expert feedback (Fig.~\ref{fig:f1scoreplot}). Figure \ref{fig:uncertainityimprovement} shows the values of uncertainty for the 5\% most uncertain instances before retraining (from \ourmethodUQ) and after retraining (from \ourframework). We can clearly notice the improvement in uncertainty scores after retraining. This result further clarifies the improved performance in \ourframework.

\begin{figure*}[!h]
    \centering
    \includegraphics[width=0.9\textwidth]{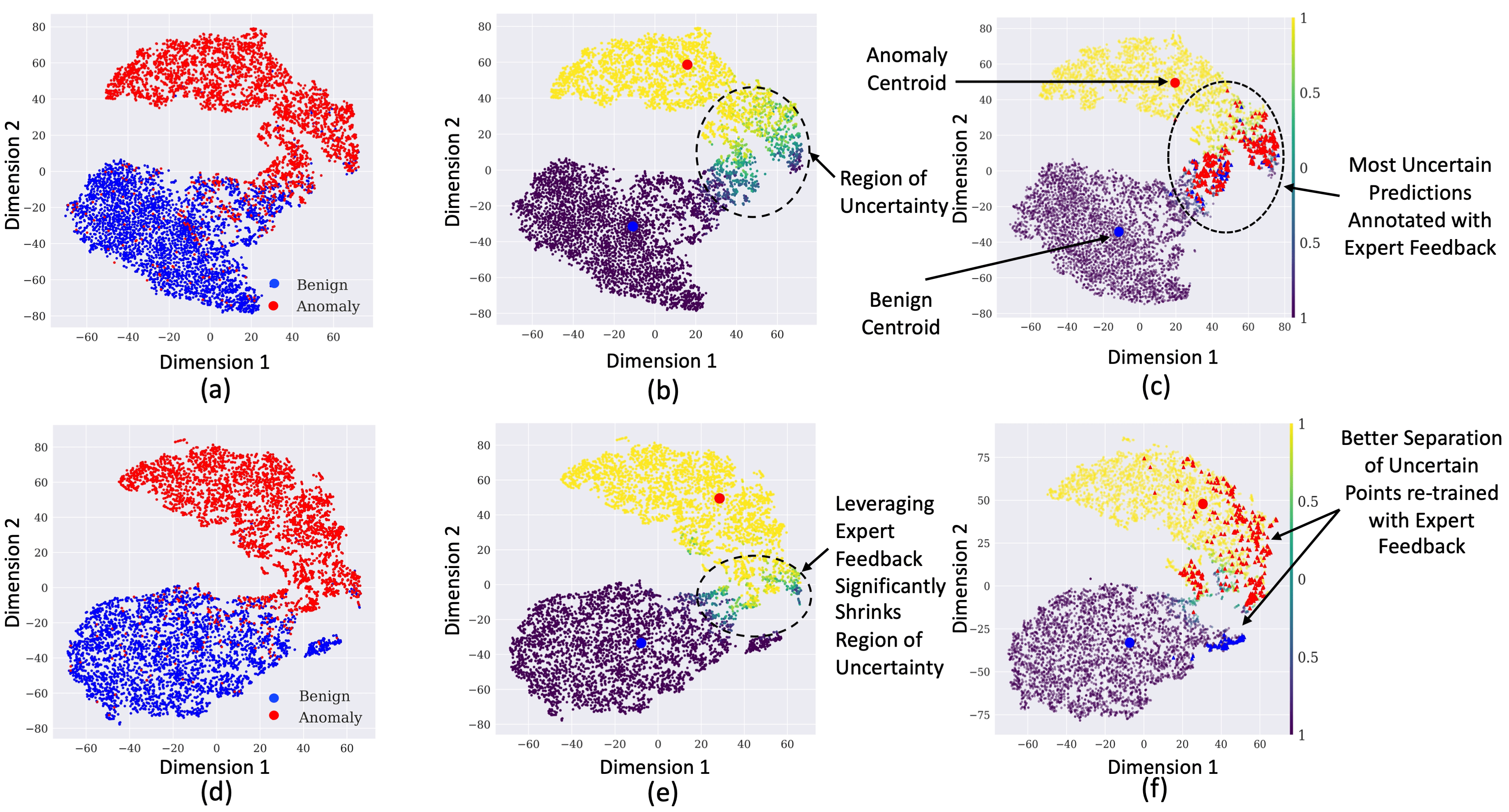}
    \caption{Figures (a) - (f) qualitatively represent the effect of incorporating human feedback in our proposed~\ourframework framework. (a) Depicts t-SNE embeddings of our~\ourmethodUQ model discriminator, colored by the ground truth labels with anomalies colored red and benign points colored blue. \hide{We immediately notice the separation of the two classes produced by the~\ourmethodUQ discriminator which is a direct effect of contrastive learning.} (b) Showcases the same t-SNE embeddings as Fig.~\ref{fig:tsneplots}a but colored by the uncertainties obtained from the~\ourmethodUQ model (yellow: anomaly with low prediction uncertainty, purple: benign with low prediction uncertainty, green regions indicate uncertain instances;). We notice a highly focused but sizeable `Region of Uncertainty' (ROU) indicated by the dotted black circle. (c) Ground truth labels of points in the ROU as specified by the expert (red: anomalous points, blue: benign points).\hide{In our experiments we solicit expert feedback for 5\% of the most uncertain points.}(d) Depicts (similar to Fig.~\ref{fig:tsneplots}a) updated t-SNE embeddings yielded by the~\ourmethodUQ discriminator after fine-tuning with expert feedback for 5\% of most uncertain instances. (e) Updated uncertainty estimates of~\ourmethodUQ post fine-tuning, we see a significant reduction in ROU (indicated by a dotted black circle) compared to Fig.~\ref{fig:tsneplots}b (f) ~\ourmethodUQ model fine-tuned with expert feedback results in greater separation between benign (blue) and anomalous instances (red) in ROU. This consequently also leads to the overall decrease in decision uncertainty as observed in Fig.~\ref{fig:tsneplots}e.}
    \label{fig:tsneplots}
\end{figure*}

\par\noindent
\textbf{Effect of Expert Feedback (Qualitative Evaluation)}: To further corroborate our claim of improved representation learning of~\ourframework due to expert feedback, we also analyze the evolution of the discriminator embeddings of~\ourframework before and after fine-tuning with expert feedback. Fig~\ref{fig:tsneplots} showcases t-SNE plots of embeddings inferred by the~\ourframework discriminator. Fig.~\ref{fig:tsneplots}a shows the representations inferred by~\ourmethodUQ before expert feedback; color indicates ground truth labels (red: anomalies, blue: benign). We notice clearly the effect of the contrastive learning employed to train the discriminator, leading to a clear separation of anomalous and benign regions in the plot. In Fig.~\ref{fig:tsneplots}b we notice~\ourmethodUQ is uncertain about a significant number of points in the inference set. This region of uncertainty (ROU - indicated by the dotted black circle in Fig.~\ref{fig:tsneplots}b) is identified and 5\% of most uncertain instances (as indicated by~\ourmethodUQ) are supplied to the expert for feedback. These expert-labeled instances are highlighted as red (expert label: anomaly) or blue (expert label: benign) points in Fig~\ref{fig:tsneplots}c. The model is fine-tuned with the full training set and the updated sets of points to produce new uncertainty estimates (Fig.~\ref{fig:tsneplots}e) wherein we see that the model is significantly less uncertain in the ROU (which has shrunk significantly). Finally, we notice that the instances that were supplied by the expert as feedback have achieved significant separation and gravitated towards their respective cluster centroids (Fig.~\ref{fig:tsneplots}f) thereby leading to improved model performance in the~\ourframework framework.

\hide{\ourmethodUQ is a sophisticated version of \ourmethod. As explained above in section ---, \ourmethodUQ provides us uncertainty values for each test instance. Figure \ref{fig:f1scoreplot} shows the trend of the F1 score as most uncertain instances are removed from inference. In this instance, we choose 5\% most uncertain instances, get opinions from domain experts and use the labels to perform finetuning using \ourmethodUQ. After fine-tuning with samples identified by domain experts using several contrastive losses, we see that in kop, the anomaly f1 score increased by 3.3\% while it increased by 5.3&\% in the 880 datasets.

Figure \ref{fig:tnseplots} shows the TNSE plots of discriminator embeddings of the test instances before and after fine-tuning. Figure a1 shows the mean monte carlo discriminator embeddings of test instances. Here anomalies are falling in one cluster and benign instances are falling in another cluster. From a2 we can see that there is a portion between the two clusters where the model has the most uncertain predictions. These are the candidate human-in-the-loop (HIL) instances. After finetuning, it's very visible in b2 that the region of uncertainty has drastically reduced. Also, the HIL instances which were picked from the most uncertain region have moved away from that region and into either of the two clusters based on if the instance is an anomaly or benign.}

\subsection{Q3. Ablation Study ~\ourframework}
\begin{table*}[ht]
    \centering
    \caption{Incremental ablation of \ourframework\\ EF: Retraining after Expert Feedback, UQ: Uncertainty Quantification and CL: Contrastive Learning.}
    \begin{tabular}{|l|C{1.3cm}|C{1.3cm}|C{1.3cm}|C{1.3cm}|C{1.3cm}|}
        \hline
        \textbf{Model} &  \textbf{Benign F1} & \textbf{Anomaly F1} & \textbf{AUROC} & \textbf{AUPRC} & \textbf{Avg.Wt. F1} \\
       \hline\hline
CAAD-EF	                        	&	\textbf{0.94}	&	\textbf{0.94}	&	\textbf{0.98}	&	\textbf{0.98}    &	\textbf{0.94}\\
\hline
CAAD-EF w/o EF	                	&	0.92	&	0.9	    &	0.97	&	0.98    &	0.91\\
\hline
CAAD-EF w/o EF, UQ	            	&	0.93	&	0.9	    &	0.97	&	0.97    &	0.92\\
\hline
CAAD-EF w/o EF, UQ, CL	            &	0.74	&	0.41	&	0.86	&	0.89    &	0.58\\
\hline
CAAD-EF w/o EF, UQ, CL, UNet		&	0.72	&	0.28	&	0.84	&	0.83    &	0.5\\
\hline
CAAD-EF w/o EF, UQ, CL, WGAN-GP		&	0.73	&	0.32	&	0.83	&	0.8	    &   0.53\\
        \hline
    \end{tabular}
    \label{tab:ablation}
\end{table*}

We have thus far verified through rigorous qualitative and quantitative experiments, the effectiveness of our proposed~\ourframework framework for AD.~\ourframework consists of multiple facets and it is important to characterize the effect of each. Hence, we conduct a detailed ablation study of the proposed~\ourframework framework.  Table~\ref{tab:ablation} details the results. We notice from the table that the~\ourframework model is dependent on each facet of its pipeline for effective representation learning with the most significant drop in performance occurring due to the removal of the contrastive learning based model training. We notice that the performance of our proposed CAAD framework is a function of the effect of contrastive learning and adversarial training. The performance is further improved with the inclusion of expert feedback (CAAD-EF). Removal of the UNet blocks from the generator also lead to deterioration in performance, primarily due to decrease in the generator learning capability. Finally, we once again notice the significant drop in performance (0.94 to 0.91 Wt. Avg. F1) when expert feedback is ignored. The results in Table~\ref{tab:ablation} further reinforce the effectiveness of~\ourframework framework for the task of AD.

%% file: sections/conclusion.tex
In this paper we have introduced~\ourmethod, a novel AD framework employing contrastive learning in an adversarial setup. We have demonstrated through rigorous experiments that our proposed method outperforms SOTA AD baselines and achieves a \textbf{92.84}\% improvement for AD in wireless communication networks as well as in more generic AD contexts. We further propose \ourframework which is a variant of \ourmethod capable of incorporating expert feedback and evaluated its effectiveness through several qualitative and quantitative experiments. Incorporating expert feedback gives a performance boost of 4.19\% over \ourmethod. Finally, we also highlight the importance of each facet of our proposed~\ourframework framework through a detailed ablation study. Moving forward we shall augment~\ourframework with more sophisticated uncertainty quantification techniques and leverage the power of our model for real-time human-in-the-loop AD applications, especially those plagued by covariate shift. The proposed models can be evaluated on more datasets to establish generalizability. We shall evaluate our methods on more sophisticated anomalies and also venture into adopting our models to detect sequential anomalies. 

%% file: sections/appendix.tex
\subsection{Wireless Dataset Description} \label{ssec:datades}
\hide{Describe in detail how the datasets were generated i.e., talk about what information packets include, how binnings is done for each dataset, etc.}
Using Software Defined Radio (SDR), we gather metadata about wireless signals, particularly FM signals. This metadata is in the form of JSON with each instance containing information about packets of signals and a timestamp. Two features we have used from this metadata are center frequency and bandwidth. We create 80x80 bins with bandwidth on the x-axis and frequency on the y-axis and populate the count of packets falling into each bin for a period of 3 minutes. This gives us a time series of 80x80 images. For context, the maximum count of packets per pixel in the \datakop dataset is 98 and the maximum count of packets per pixel in the \datamltest dataset is 201. Also, the mean of maximum count of packets per pixel per image is 20.5 for \datakop and 107.1 for \datamltest. This indicates that \datamltest is dense compared to \datakop. Min-max normalization based on the global min and max of the count of packets per pixel of the training set is performed.  Once such normalized images are gathered, we denoise the training set by masking out pixels that have a probability of having a non-zero value of $<$ 0.0005.\\
\subsection{Types of Wireless Anomalies}

\begin{figure}[!h]
    \includegraphics[width=0.98\columnwidth]{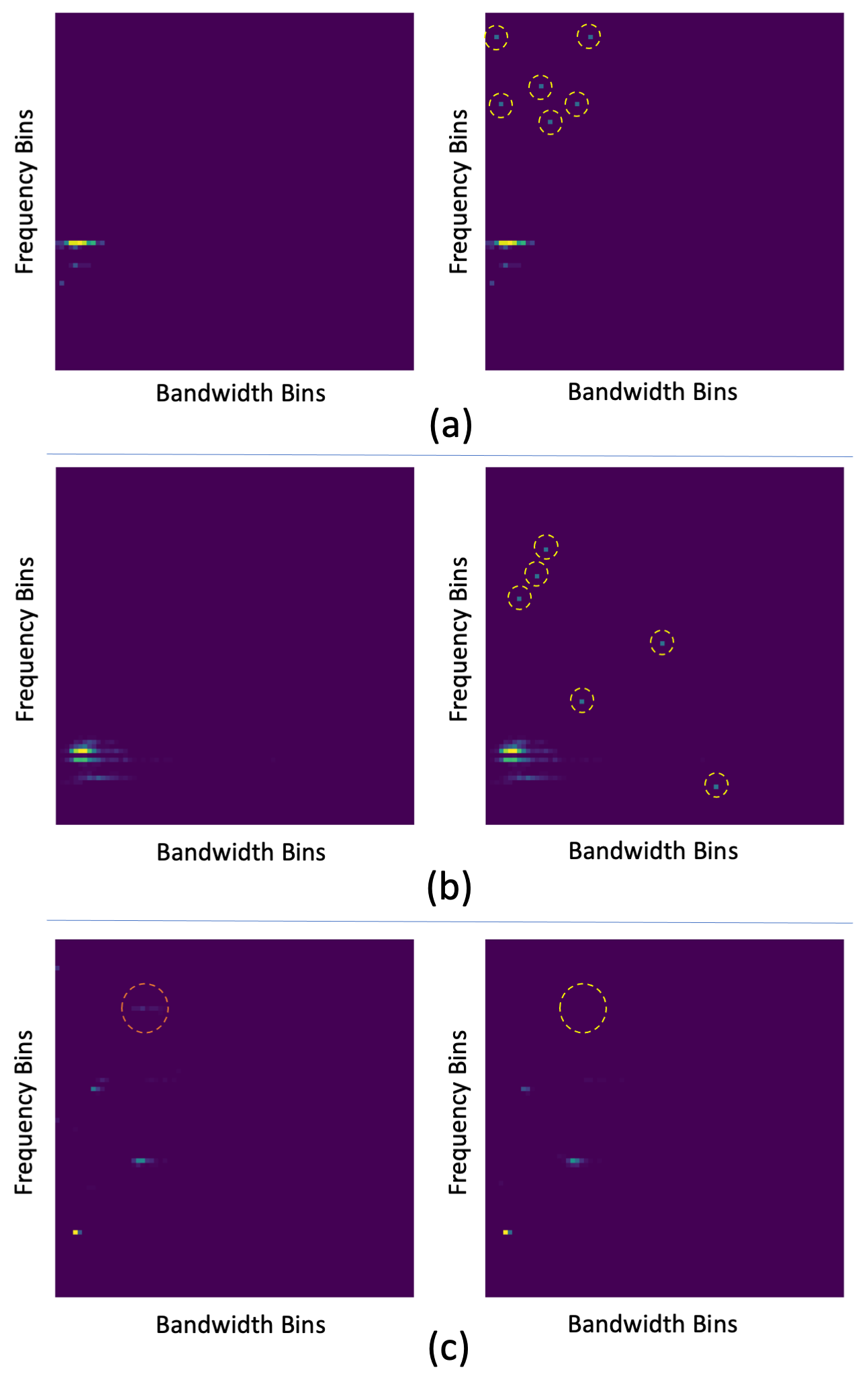}
    \caption{These are examples of preprocessed wireless emission activity data that we input into our models. The left figures indicate benign instances and the right figures indicate corresponding anomalous instances. Anomalies are annotated with dashed yellow circles. Figure (a) is a sample from the LTW1 dataset. Anomalies in the right figure of (a) are 7x amplified in order to improve visibility. Figure (b) is a sample from the LTW2 dataset. Anomalies in the right figure of (b) are 30x amplified in order to improve visibility. Figure (c) is a sample from the STW1 dataset. In the left figure in (c), the orange circle points to the signal that goes missing in the right figure (Drop anomaly).}
    \label{fig:wirelesssamples}
\end{figure}

\begin{enumerate}
    \item \textbf{Hopper Anomalies:} In this type of anomaly, frequency hopper signals are transmitted. This creates activity in different regions of the 80x80 image. Examples of such hopper anomalies are shown in Figures \ref{fig:wirelesssamples}a and \ref{fig:wirelesssamples}b.
    \item \textbf{Drop Anomalies:} In this type of anomaly, an LTE signal in the cellular band goes offline at around 198 seconds into the dataset. We use until 170 seconds of data from training and the rest, with anomalies for testing. This removes a region of activity from the 80x80 image. Examples of a drop anomaly  is shown in Figure \ref{fig:wirelesssamples}c.
    \item \textbf{Naturally occurring anomalies:} In the training set, as mentioned in Section \ref{ssec:datades}, we mask out pixels that have a probability of having a non-zero value of $<$ 0.0001. However, we do not do this in the testing set and we instead label such images as anomalies.
\end{enumerate}
\subsection{Anomaly Injection}
\hide{Talk about the random injection of anomalies. Second also talk about how the test set noise was also included as anomalies based on some threshold.}
\begin{enumerate}
    \item \textbf{Injection of hopper anomalies:} The available number of real-world hopper anomalies was 60 packets with different bandwidths and frequencies for each packet. These 60 packets were then grouped into numbers of 6 which effectively produced 10 different anomaly signatures. Now, every anomaly signature contains 6 anomalies with each having a count (density) of 1. We use these anomaly signatures as template anomalies and inject (add) them in test instances without anomalies while also keeping a copy of the original test instance without anomalies. Before anomaly injection, the template is min-max normalized using the global min and max of the count of packets per pixel of the entire training set. As \datamltest has max of the count of packets per pixel of the entire training set value as 201 and since our anomalies have a density of 1, during min-max normalization, the injected anomalies become very subtle. This is one explanation for the relatively low anomaly F1 scores in \datamltest compared to other datasets. Figure \ref{fig:wirelesssamples}b shows anomalies amplified by 30x in order to be visible to the naked eye.
    \item \textbf{Injection of Drop Anomalies:} To inject drop anomalies, the LTE signal is manually turned off.

\end{enumerate}
\hide{
\subsection{MNIST Anomaly}
Describe the MNIST Anomaly injection process.
In the MNIST dataset, we consider the digit '4' as benign and all other digits as anomalous. We train our models with the images of '4', and expect our models to identify other digits as anomalies. By nature, there are an equal number of instances for each digit in the MNIST test set and hence in our test set, we have 9 times the number of benign as the total number of anomalies.}